\documentclass[sigconf, 9pt]{acmart}
\AtBeginDocument{%
  }
\newcommand{\name}{GazeSummary}
\newcommand{\sssec}[1]{\vspace{0.0in}\noindent\textbf{#1}}
\usepackage{subcaption}
\usepackage{graphicx}
\usepackage{multirow}
\usepackage{booktabs}
\usepackage{xcolor}     
\usepackage{pifont}

\definecolor{c_density}{RGB}{114, 182, 161}
\definecolor{c_heatmap}{RGB}{233, 150, 117}
\definecolor{c_svm}{RGB}{149, 163, 195}
\definecolor{c_text}{RGB}{219, 150, 192}
\definecolor{c_target_para}{RGB}{162, 200, 101}
\definecolor{c_user}{RGB}{229, 201, 73}

\definecolor{c_density}{RGB}{34, 102, 81}
\definecolor{c_heatmap}{RGB}{153, 70, 37}
\definecolor{c_svm}{RGB}{69, 83, 115}
\definecolor{c_text}{RGB}{139, 70, 112}
\definecolor{c_target_para}{RGB}{82, 120, 21}
\definecolor{c_user}{RGB}{159, 131, 3}
\copyrightyear{2026}
\acmYear{2026}
\setcopyright{cc}
\setcctype{by}
\acmConference[HotMobile '26]{The 27th International Workshop on Mobile Computing Systems and Applications}{February 25--26, 2026}{Atlanta, GA, USA}
\acmBooktitle{The 27th International Workshop on Mobile Computing Systems and Applications (HotMobile '26), February 25--26, 2026, Atlanta, GA, USA}
\acmPrice{}
\acmDOI{10.1145/3789514.3792037}
\acmISBN{979-8-4007-2471-8/2026/02}

\begin{document}

\title{GazeSummary: Exploring Gaze as an Implicit Prompt for Personalization in Text-based LLM Tasks}



\author{Jiexin Ding}
\affiliation{%
  \institution{University of Washington}
  \city{Seattle}
  \country{USA}}
\email{jxding@uw.edu}

\author{Yizhuo Zhang}
\affiliation{%
  \institution{University of Washington}
  \city{Seattle}
  \country{USA}}
\email{mattyz@uw.edu}

\author{Xinyun Liu}
\affiliation{%
  \institution{Google}
  \city{Seattle}
  \country{USA}}
\email{xinyunliu@google.com}

\author{Ke Chen}
\affiliation{%
  \institution{University of Washington}
  \city{Seattle}
  \country{USA}}
\email{katechen@uw.edu}

\author{Yuntao Wang}
\affiliation{%
  \institution{Tsinghua University}
  \city{Beijing}
  \country{China}}
\email{yuntaowang@tsinghua.edu.cn}

\author{Shwetak Patel}
\affiliation{%
  \institution{University of Washington}
  \city{Seattle}
  \country{USA}}
\email{shwetak@cs.washington.edu}

\author{Akshay Gadre}
\affiliation{%
  \institution{University of Washington}
  \city{Seattle}
  \country{USA}}
\email{gadre@uw.edu}

 \renewcommand{\shortauthors}{Ding et al.}

\begin{abstract}

Smart glasses are accelerating progress toward more seamless and personalized LLM-based assistance by integrating multimodal inputs. Yet, these inputs rely on obtrusive explicit prompts. The advent of gaze tracking on smart devices offers a unique opportunity to extract implicit user intent for personalization. 
This paper investigates whether LLMs can interpret user gaze for text-based tasks. We evaluate different gaze representations for personalization and validate their effectiveness in realistic reading tasks. Results show that LLMs can leverage gaze to generate high-quality personalized summaries and support users in downstream tasks, highlighting the feasibility and value of gaze-driven personalization for future mobile and wearable LLM applications.

\end{abstract}

\maketitle

\section{Introduction}\label{sec:intro}


Smart glasses with see-through displays and AI assistants are bringing in a new era of mobile computing. Beyond traditional metrics like throughput, latency, and battery life, these devices increasingly focus on delivering personalized user experiences. Current designs rely on explicit prompts via microphones paired with LLMs such as ChatGPT, Gemini, and Grok. While effective, these prompts require user effort and are often cumbersome or impractical. For example, a user attending a research talk would need to repeatedly say ``this slide is important'' to guide a summarization system, which is intrusive and unnatural.

Gaze tracking is rapidly becoming standard in XR headsets and is widely expected to be included in the next generation of smart glasses (Project Aria), making it a practical implicit input for LLM-based tasks.
This paper focuses specifically on the above task of generating a personalized summary based on user gaze. 
Prior work has shown that gaze can effectively reflect user attention~\cite{cui_gaze_2024} and can support extractive summarization through gaze density~\cite{Vazaios_2023_eye_summary}, classifiers~\cite{Taieb-Maimon01092024}, or pre-trained models~\cite{sunlifeng_2020, dubey_2020_wikigaze}.
However, no existing work integrates gaze directly with LLMs leveraging their strengths in text-based tasks.



\begin{figure}[t]
  \centering
  \includegraphics[width=1.0\linewidth]{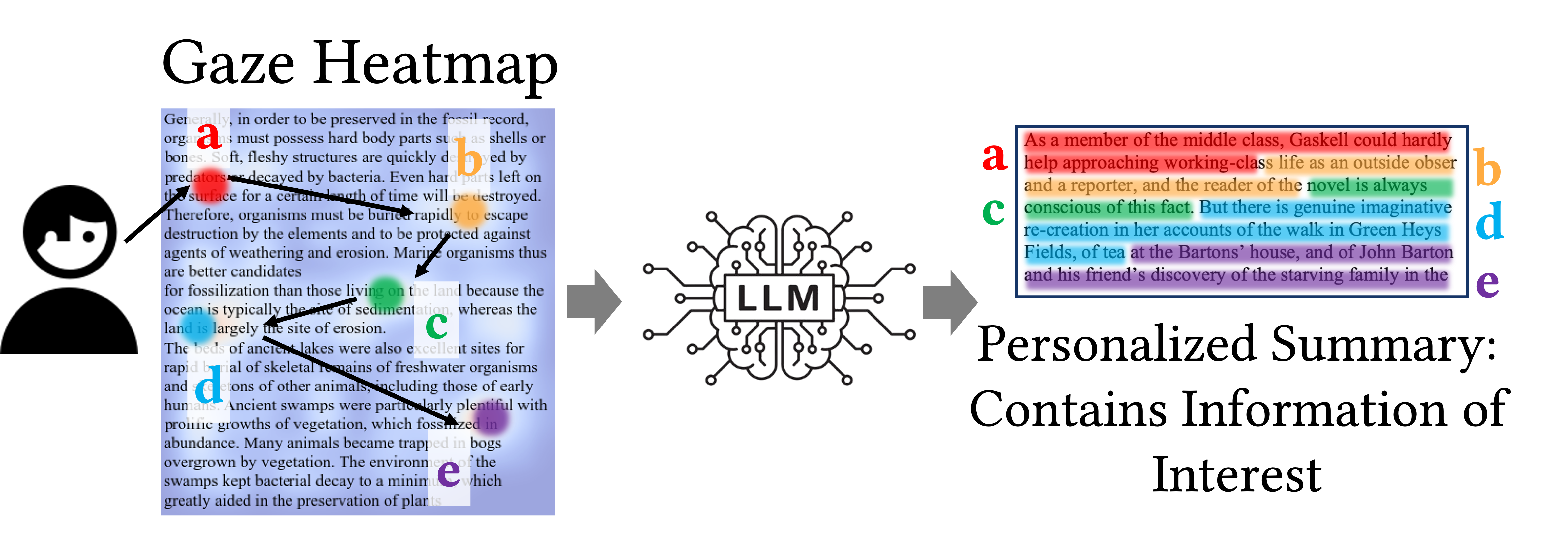}
  \caption{By integrating gaze as a proxy for user attention, our approach enables large language models to produce summaries centered on the content users find most relevant to their interest.}
  \label{fig:intro}
  \vspace{-0.2in}
\end{figure}

This paper, \name, examines how LLMs can interpret gaze data to generate summaries that align with users’ attention. While prior work shows it is feasible to personalize summary using gaze, most approaches use gaze only to locate relevant text and have not explored how to integrate gaze directly into LLMs and \textbf{what is the best representation of gaze for such integration}. We address this gap by evaluating multiple gaze representations, finding that gaze heatmap performs best in its similarity with user-provided summaries. It captures fine-grained attention patterns thus outperform explicit focus inputs, particularly at the phrase level.

We further explore the potential of gaze-based personalization in enhancing real-world user experience and productivity by developing a prototype gaze-guided assistant for academic reading. It automatically generates personalized summaries for users based on gaze heatmaps. We conducted \textbf{a user study} with 10 participants \textbf{that compares the effectiveness of \name\ with traditional methods}, such as summarization based on explicit user prompts and text-only input, in creation-oriented tasks like writing. The results show that \name\ outperforms text-only summarization and achieves comparable performance to user-prompt-based summarization, while requiring lower mental effort to use.

The main takeaways of this work are as follows:
\begin{itemize}
    \item We explore several gaze representations as the input for LLM for personalized text summarization and find heatmap outperforms others and even the explicit-prompt-based summary at phrase level. 
    \item We build a gaze-guided assistant for academic reading and conduct a user study which shows that \name\ is more helpful and easy-to-use compared to the commonly used summarization methods.
\end{itemize}

\section{Related Work}\label{sec:relatedwork}



\sssec{Gaze Tracking and LLMs:} With the rapid development of large language models (LLMs), recent research has begun exploring ways to integrate gaze signals into LLMs. Some works like GazeNoter~\cite{tsai2024gazenoter} and EmBARDiment~\cite{Bovo_2025_emBARDiment} treats gaze purely for selecting outputs or objects of importance, and incorporates this information into the LLM prompt, resulting in more efficient and context-aware interactions. Other works have treated gaze as a visual modality and input it into vision-language models (VLMs) to utilize the models’ visual encoder to interpret gaze. By aligning gaze with the real world, these approaches~\cite{wang2024gvoila,zeyu2024Voila-A,lee2025walkie} achieve more efficient visual question answering (VQA) solutions that incorporate image features with queries to respond to the user. While these approaches have combined gaze and LLMs for interaction and VQA tasks, text-based scenarios are very different: gaze data is much denser, and text-related tasks require finer granularity, often at the word level. To the best of our knowledge, no work has systematically investigated optimum gaze representations for LLMs to understand and leverage in text-related tasks, such as summarization.

\sssec{Personalized Text Summarization:}
Summaries condense information to highlight key ideas, enhancing understanding and retention~\cite{chiu2013integrating}. However, generic summaries often fail to match what individual users consider important, highlighting the need for personalized summaries~\cite{bhatnagar2023personalized}. Previous works on personalization determine user context based on the background information~\cite{kirstein-etal-2024-tell, ao2021pens, zhang2024personalsum} or users' explicit feedback~\cite{Ghodratnama2020Adaptive}. 
However, these approaches mainly rely on past information about the user and thus cannot dynamically adapt summaries to reflect the user's current reading state and focus.

Given the strong links between gaze patterns and cognitive states~\cite{southwell2023gaze,Rayner2009The3S,just2018using}, some works use gaze to detect content of interest during reading and compile it into personalized summaries. Research shows that re-reading important parts and spending more time on them improves comprehension, and are reflected in gaze data~\cite{southwell2023gaze,ikhwantri-etal-2024-analyzing}. Based on this, some methods generate personalized summaries by selecting the most focused sentences or UIs according to gaze density~\cite{Vazaios_2023_eye_summary, dubey_2020_wikigaze, sunlifeng_2020}, while others aggregate gaze into features such as fixations to identify key sentences~\cite{Taieb-Maimon01092024}. However, most of these approaches simply concatenate key sentences or use basic word embeddings, resulting in summaries that are less fluent and coherent than generic ones.

\section{Gaze Representations}\label{sec:gaze_representation}
This section explores the best representation of gaze information for text-based LLM tasks using traditional encoding methods. Currently, there is no consensus on the optimal way to encode gaze features via LLM prompts for summarization tasks. Existing approaches vary significantly in terms of the type and amount of gaze information preserved, as well as how explicitly that information is encoded within the model input.

We identify two key dimensions along which gaze representations differ: (1) \textit{\textbf{Information richness}}: the extent to which raw or derived gaze features (e.g., fixation duration, saccade pattern, regressions) are preserved. (2) \textit{\textbf{Visual explicitness}}: the degree to which gaze signals are encoded in a way that is easily interpretable by models. Based on these criteria, we analyzed existing gaze representation methods and identified three representative strategies with distinct trade-offs: \textit{gaze density}, \textit{gaze heatmap}, and \textit{attention-based gaze (via SVM)}. These methods span different points along the two axes: gaze density is simple and visually salient but low in informational content; heatmaps encode rich attention distributions but in a form that is harder for the model to interpret; attention-based gaze modeling provides a more balanced representation with moderate interpretability and feature-level abstraction. We use Gemini 2.5 Pro model for summary generation known to be one of the best LLMs with visual ability. The summary is set to be one-fifth the length of the original text. The methods we explore are as follows:

\begin{figure}[t]
  \centering
  \includegraphics[width=1.0\linewidth]{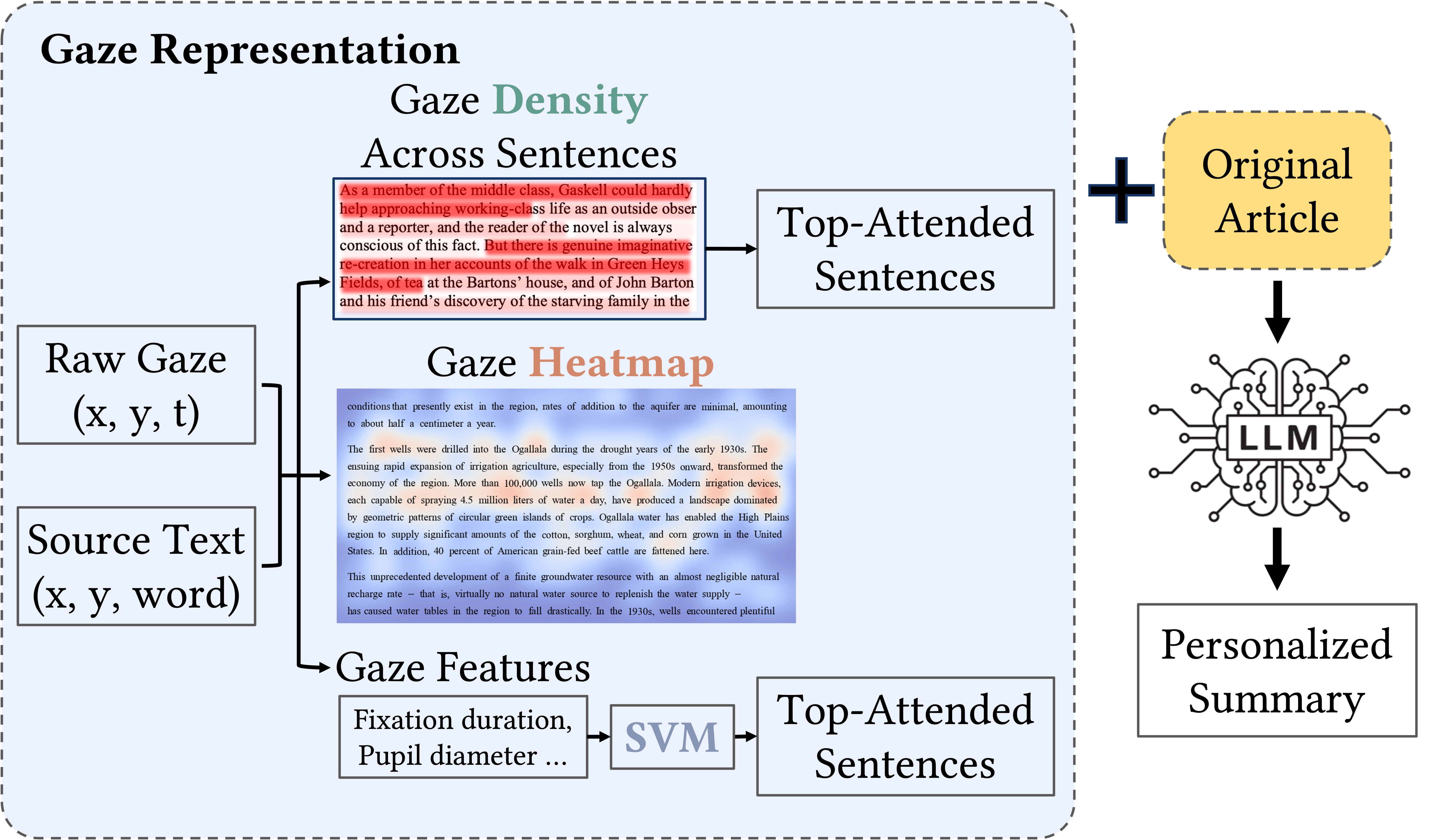}
  \caption{Generate personalized summary using different gaze representations.}
  \label{fig:gaze_repre}
  \vspace{-0.2in}
\end{figure}

\sssec{\underline{Gaze \textcolor{c_density}{Density}}:} The gaze density method identifies sentences a reader focuses on most by computing total fixation duration per sentence. Sentences are first segmented, and gaze data -- aligned via word bounding boxes -- is aggregated over all tokens. Sentences are then ranked by gaze duration, and the top 20\% (typically 5-7) are selected as personalized focus content. These high-attention sentences are inserted into a text-based prompt to generate summary.

\sssec{\underline{Gaze \textcolor{c_heatmap}{Heatmap}}:} Gaze heatmaps are generated by aggregating gaze points over the reading area and overlaying them on the text, with warmer colors indicating higher fixation density. The final output is a multimodal image-text pair, where the heatmap is presented alongside the original text as visual context for the LLM. Because heatmaps are implicit, prompt engineering is needed to convey gaze salience. Pilot experiments show that including color explanations and step-wise guidance improves content relevance and coherence.

\sssec{\underline{Gaze Attention \textcolor{c_svm}{SVM}:}} For the feature engineering, we adopt a sliding window approach by dividing the reading session into 4-second intervals. For each window, we extracted 26 gaze features using PyGazeAnalyser\footnote{\url{https://github.com/esdalmaijer/PyGazeAnalyser/blob/master/pygazeanalyser/detectors.py}}, including fixation duration, saccade metrics (duration, angle), and pupil diameter according to~\cite{bixler_automatic_2016}. We used dataset from our previous work~\cite{unknown_2025_ding} as the training data as it labels words based on the time spent by the user on that word. We label a window as ``high-attention'' if it contains at least one words that labeled as ``focused'' by the participants. Otherwise, it is labeled as 0. We used SVM and performed grid search to find the best parameters ($class\_weight=`balanced', C=1, gamma=0.01$). The accuracy is 65.59\% on our dataset which is slightly lower the the result reported in~\cite{bixler_automatic_2016}. During the inference, the sentence is labeled as ``high-attention'' if at least one of the window that covers this sentence is labeled as ``high-attention''. Then we ranked these sentences by confidence score and put top 20\% sentences into the prompt.

\section{Evaluation of Gaze Representations}

\label{sec:gaze_evaluation}
We compare the above three gaze representation designs  (\textcolor{c_density}{Density}, \textcolor{c_heatmap}{Heatmap}, \textcolor{c_svm}{SVM}) with two baselines: a text-only method (\textcolor{c_text}{Text}) and a explicit-prompt method with target paragraphs included in the prompt (\textcolor{c_target_para}{Target Paragraphs}). We evaluate these gaze representations for two metrics:
\begin{enumerate}
    \item Efficiency for enabling personalization of summary 
    \item Overall quality of the generated summary 
\end{enumerate}

\begin{figure}[h]
    \centering
    \includegraphics[width=1.0\linewidth]{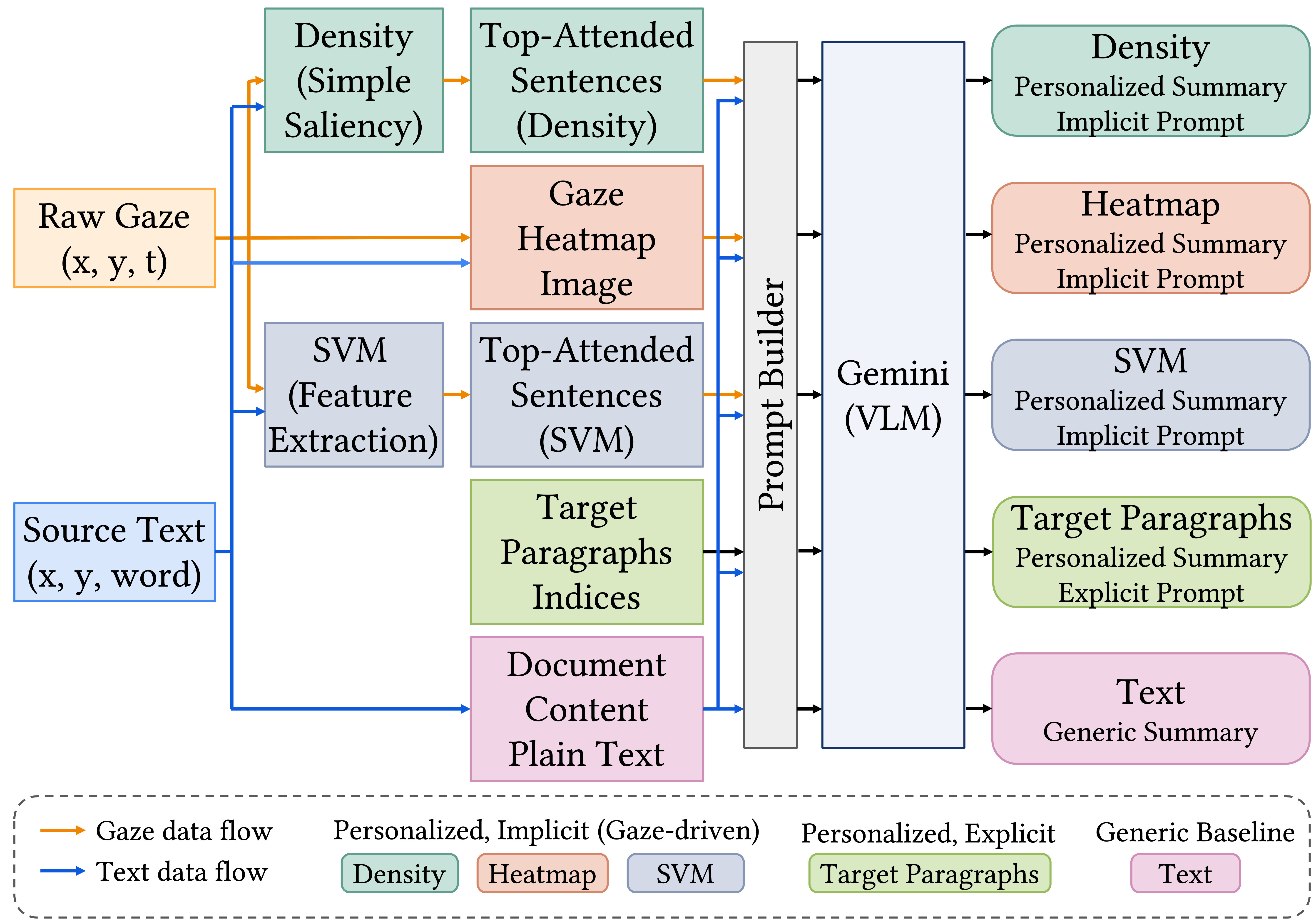}
    \vspace*{-0.25in}
    \caption{Implementation of different summarization methods. Refer to Appendix A for the details of prompts.}
    \vspace*{-0.2in}
    \label{fig:methods_implementation}
\end{figure}

\noindent Fig.~\ref{fig:methods_implementation} presents the implementation of each method
, while Table~\ref{tab:methods_comparsion} summarizes their distinctions. The comparison among the three gaze-based methods identifies the most effective gaze representation. Contrasting these methods with Target Paragraphs reveals the relative strengths of implicit versus explicit prompting. Comparing all personalized methods with the Text baseline further demonstrates the advantages of personalized summaries over generic (non-personalized) ones.

\begin{table}[h]
\begin{tabular}{lll}
\toprule
                  & Personalization & Prompt Type               \\    \midrule
Density           & {\color{green}\ding{51}} & Implicit        \\
Heatmap           & {\color{green}\ding{51}} & Implicit        \\
SVM               & {\color{green}\ding{51}} & Implicit       \\
Target Paragraphs & {\color{green}\ding{51}} & Explicit \\
Text              & {\color{red}\ding{55}} & Not applicable              \\ \bottomrule
\end{tabular}
\caption{Comparison among different summarization methods.}
\label{tab:methods_comparsion}
\vspace*{-0.2in}
\end{table}

\subsection{User Study Design}
\label{sec:user_study}

A user study was conducted to answer these two questions. A total of 10 participants took part in the experiment, consisting of 4 males and 6 females, with an average age of 25.4 years.  In the user study, we collected user gaze data with attention variation and summaries written by users. Users were informed of two topics (corresponding to two target paragraphs) to simulate the scenario of users reading an article with a purpose. This study follows a within-subject experimental design where each participant completes a structured reading task across two TOEFL reading passages. The experimental workflow consists of three primary phases: calibration, reading and writing summary. 

Before reading, participants completed a standard gaze calibration using the Tobii Pro Spark eye tracker. They were then presented with two randomly selected TOEFL reading passages (700–800 words each) with similar difficulty and length. Then, they were asked to study on two topics that are corresponding to two paragraphs prior to reading to guide attention. After reading each passage, participants wrote a personalized summary of 100–150 words focusing on the main idea and the two assigned topics. These user-provided summaries will be used to validate our analysis later.




\begin{figure*}[t]    
  \centering
  \includegraphics[width=\textwidth]{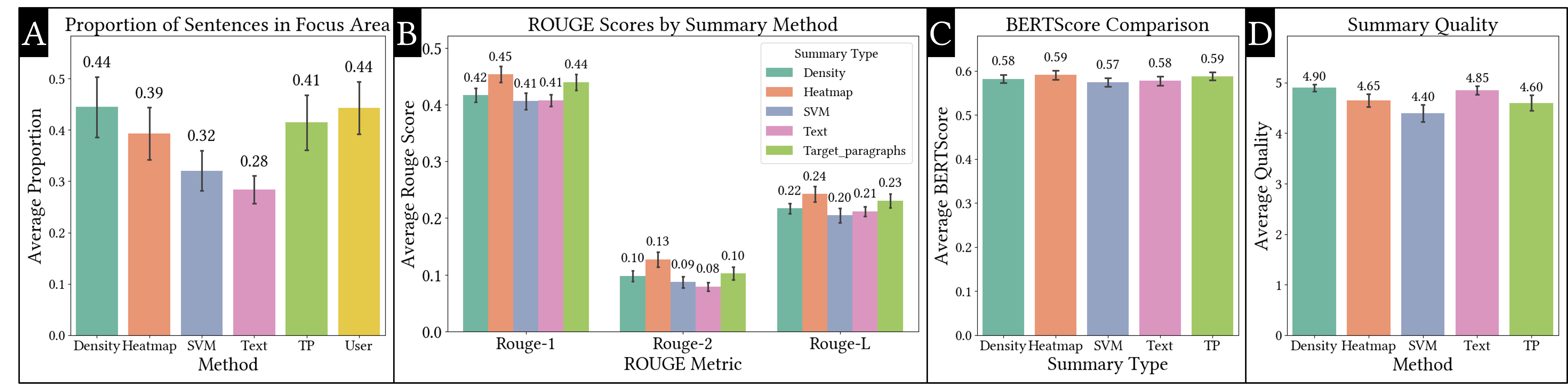}
   \vspace*{-0.25in}
  \caption{(A) Gaze effectively guides LLMs to generate summaries that focus on the content users attend to. (B) At the lexical level, \textcolor{c_heatmap}{Heatmap} outperforms other gaze representations and even surpasses explicit-input summaries (\textcolor{c_target_para}{Target Paragraphs}) at the phrase level. (C) Semantically, there is no significant difference. (D) Personalization does not significantly degrade summary quality.}
 \vspace*{-0.2in}
  \label{fig:result}
\end{figure*}

\subsection{Results}
\sssec{Content Focus Analysis:} We measure the summary focus by calculating the proportion of sentences in the summary that corresponded to these target paragraphs in the \textbf{source text}. Given that the two articles contained six paragraphs, a uniformly distributed summary (Text) would have a target-related sentence ratio of $\sim$0.317. A significantly higher ratio indicates that the summary content is more aligned with the target paragraphs, effectively capturing the gaze information.

We compute BERTScore~\cite{zhang2019bertscore} between each summary sentence and all paragraphs in the original text to identify its most semantically related paragraph. 
As shown in Fig.~\ref{fig:result}(A), the summary provided by the user and Density have the highest scores, followed by the Target Paragraphs and Heatmap. 
The reason why Density performs better than Heatmap and Target Paragraphs based on this metric is that this metric only considers sentence-level similarity. Density directly reflects this sentence-level user's focus and does not require LLM to perform additional reasoning, while Heatmap contains more global information and Target Paragraphs reflects more coarse-grained information.
Their overlap ratios were significantly higher than Text ($p < 0.05$), demonstrating the value of gaze data for producing personalized user-focused summaries. SVM showed no significant improvement over Text, likely due to low attention-detection accuracy. Overall, gaze information enables LLMs to better capture user focus and generate personalized summaries.

\label{sec:similarity}
\sssec{Similarity with the User-Provided Summary:}  We analyze the overlap of gaze-based summary with \textbf{user-provided summary} to assess overlap with user focus. Each participant was asked to write a personalized summary after reading each passage. We quantify lexical semantic similarity and keyword-level comparison between the user summary and the five generated summaries using ROUGE scores (ROUGE-1, ROUGE-2, and ROUGE-L~\cite{lin2004rouge}) and BERTScore. Higher similarity for the gaze-based summary indicates greater alignment with the user’s retained and emphasized content. The Wilcoxon signed-rank test is performed to compare the three gaze-based summaries (Density, Heatmap, SVM) and the text-only summary (Text) and explicit-prompt summary (Target Paragraphs). The $p-value$ is shown in Table~\ref{tab:similarity_significance}.

Fig.~\ref{fig:result}(B) demonstrates Heatmap is significantly better than Text  under all metrics. Density is also significant better than Text under ROUGE-2. When compared to summary generated by explicit input (Target Paragraph), Heatmap is also significantly better ($p = 1.97e-3$) under ROUGE-2. This result indicates that the heatmap is the most effective gaze representation, performing \textbf{even better} than directly providing the LLM with the target paragraphs at the phrase level. This is due to the fact that Heatmap summary conveys sub-sentence-level information, such as which specific words or phrases the user fixated on, offering a level of granularity that cannot be achieved even when the user explicitly specifies their interests.  Fig.~\ref{fig:result}(C)  also shows that there was no significant semantic difference among the summaries, although the Heatmap achieved slightly higher scores.



\sssec{Overall Performance Rated by LLM:}\label{sec:gpt_rating}
Previous non-LLM-based personalized summarization approaches often experience a decline in overall summary quality. We compare the overall quality of gaze-based summaries to text-only summary by leveraging the LLM itself to rate the machine-generated summaries. 
Following protocols from recent literature~\cite{shakil2024evaluating}, LLMs were prompted with both the source text of the original article and the candidate summary, and asked to assign 1-5 score for the quality of summary considering 4 aspects: Consistency, Coherence, Relevance and Fluency~\cite{fabbri2021summeval}. The model used for this task is Gemini 2.5 Pro.

\begin{table}[t]
\begin{tabular}{lllll}
\toprule
Gaze sum. & Rouge-1          & Rouge-2          & Rouge-L          & Bertscore \\ \midrule
Density     & 0.47             & \textbf{4.84e-2} & 0.96             & 0.67      \\
Heatmap     & \textbf{3.65e-3} & \textbf{1.68e-4} & \textbf{2.66e-2} & 0.06      \\
SVM         & 0.87             & 0.37             & 0.28             & 0.52      \\ 
\bottomrule
\end{tabular}
\caption{Comparison ($p$-$value$) between gaze-based summaries and the text-only summary. Bold indicates significant difference ($p < 0.05$)}
\label{tab:similarity_significance}
\vspace*{-0.4in}
\end{table}

Fig.~\ref{fig:result}(D) shows that, except for the SVM condition, no significant differences in overall summary quality were observed among the methods. These minor quality differences may result from the varying reasoning complexity of the methods. For example, Heatmap performs slightly worse than Density and Text, likely due to the added complexity of multimodal inputs, although the difference is not statistically significant. The lower score of Target Paragraphs compared to Text suggests that focusing content more narrowly on high-attention paragraphs may come at a minor cost to overall quality. However, this difference was not substantial, indicating that incorporating gaze information can make summaries more personalized without compromising their overall quality.


\vspace{0.0in}\noindent\fbox{\parbox[][][]{0.97\linewidth}{\textbf{Takeaway 1:} \textit{Heatmap is an ideal gaze representation for text-based LLM-assisted tasks outperforming explicit user input at sub-sentence level while maintaining user focus, semantic consistency and summary quality.}}}

\section{Case Study -- Gaze-Guided Assistant for Academic Reading}
Considering the phrase-level advantages of heatmap-based gaze representation encoding in text-based LLM tasks, we develop a prototype academic reading application, since academic articles often contain many multi-word expressions and fixed phrases. The system automatically generates summaries after reading and displays them as references during writing. In practical scenarios users must process large volumes of literature, where these summaries could serve as a cognitive offloading tool, supporting learning and writing.

\subsection{Evaluation Metrics for Summary}
We aim to investigate whether \name\ can provide tangible benefits for users' reading and writing. Participants rated each method on four dimensions: (1) \textit{Perceived Usefulness}, (2) \textit{Information Coverage}, (3) \textit{Confidence and Trust}, and \textit{(4) tool effort}. They were also asked to rank the overall preference on a 5-point Likert scale as follows:

\sssec{Perceived Usefulness:} How helpful was the summary for completing the task? \textbf{1}-\textit{Not helpful}, \textbf{5}-\textit{Extremely helpful}.

\sssec{Information Coverage:} How many information points did the summary cover that you focused on while reading the original article? \textbf{1}-\textit{Did not cover}, \textbf{5}-\textit{Covered perfectly}.

\sssec{Confidence and Trust:} How confident are you that the summary accurately reflects the core facts and data from the original article? \textbf{1}-\textit{Not confident}, \textbf{5}-\textit{Extremely confident}.

\sssec{Tool Effort:} Thinking about the process of generating summary, how much effort did this specific summarization process require from you? \textbf{1}-\textit{Little effort}, \textbf{5}-\textit{Great effort}.

We compare \name\ with two baselines described previously : Text and Target Paragraphs. Comparing \name\ with Text allows us to assess the effect of personalization on user performance. On the other hand, \name\ with Target Paragraphs tests whether implicit input via gaze offers comparable performance to explicit input.

\subsection{User Study Design}

\begin{figure}[t]
  \centering
  \includegraphics[width=1.0\linewidth]{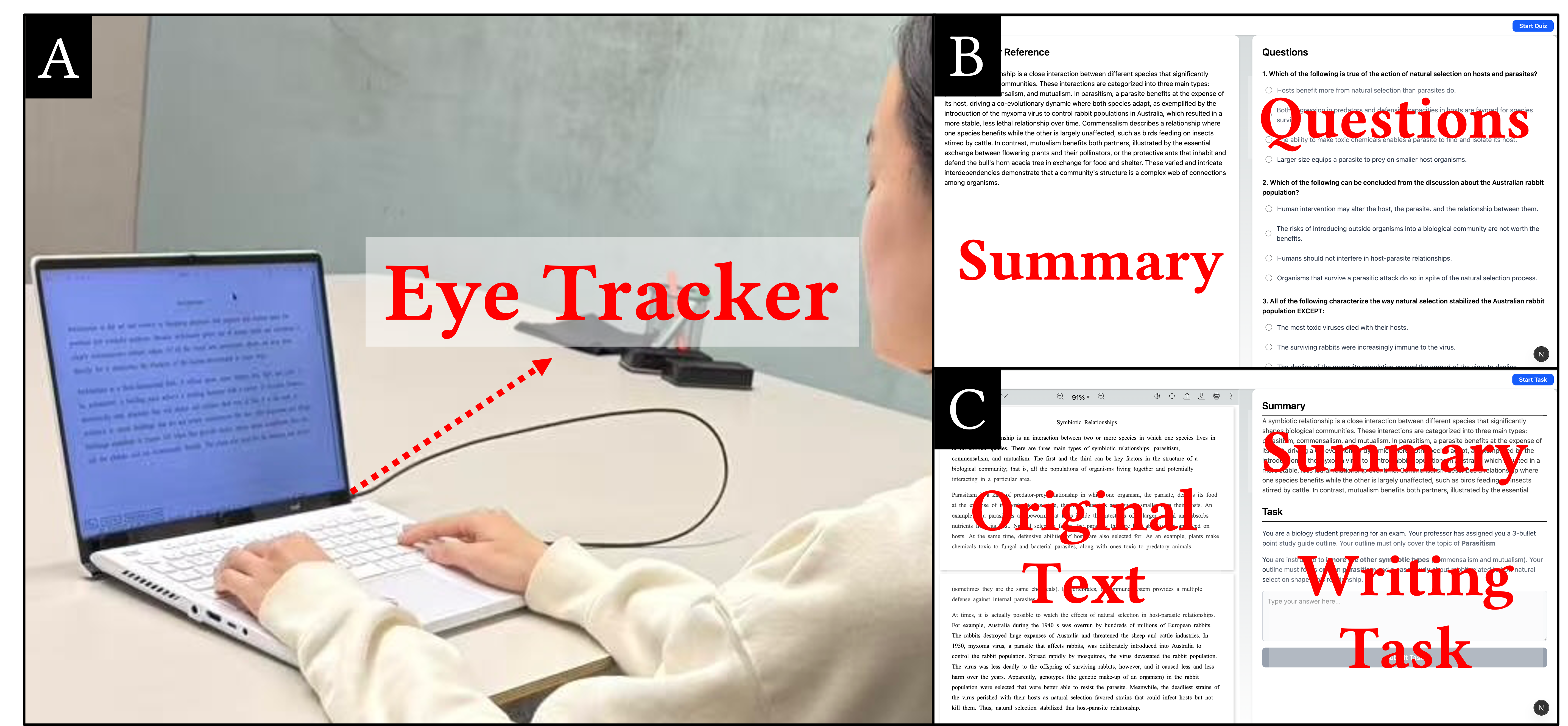}
  \caption{(A) During the user study, the eye tracker was fixed below the screen. (B) User interface for questions. (C) User interface for writing task.}
  \label{fig:user_study}
\end{figure}

The experiment
consisted of three phases: eye-tracker calibration, reading and task completion, and questionnaire. Each participant read three TOEFL passages ($\sim$700 words, six paragraphs each), with a different summarization method applied to each passage. The order of methods was randomized and not disclosed to participants.

Before reading each passage, participants were shown a brief description of a writing task, including 
a topic corresponding to two consecutive paragraphs in the article, and instructions on which content to ignore. After memorizing the task, participants read the passage for three minutes (Fig.~\ref{fig:user_study}(A)).
This setup encourages selective reading and induces gaze distributions, simulating typical reading scenarios where attention is guided by goals.

After reading, participants answered three reading comprehension questions on the assigned topic (Fig.~\ref{fig:user_study}(B)). They could refer to the summaries but not the original text to ensure careful engagement with the article and summary. Then, participants wrote a three-bullet-point note as the writing task (Fig.~\ref{fig:user_study}(C)). They were allowed to refer both the summary and the article, simulating a realistic writing scenario. Upon task completion, participants filled out a questionnaire.

We recruited 10 participants, including two native speakers and eight non-native speakers; six wore glasses and four did not. The eye-tracker, LLM model, and prompts were identical to those described in Section 4.

\subsection{Results}
\begin{figure}[t]
  \centering
  \includegraphics[width=1.0\linewidth]{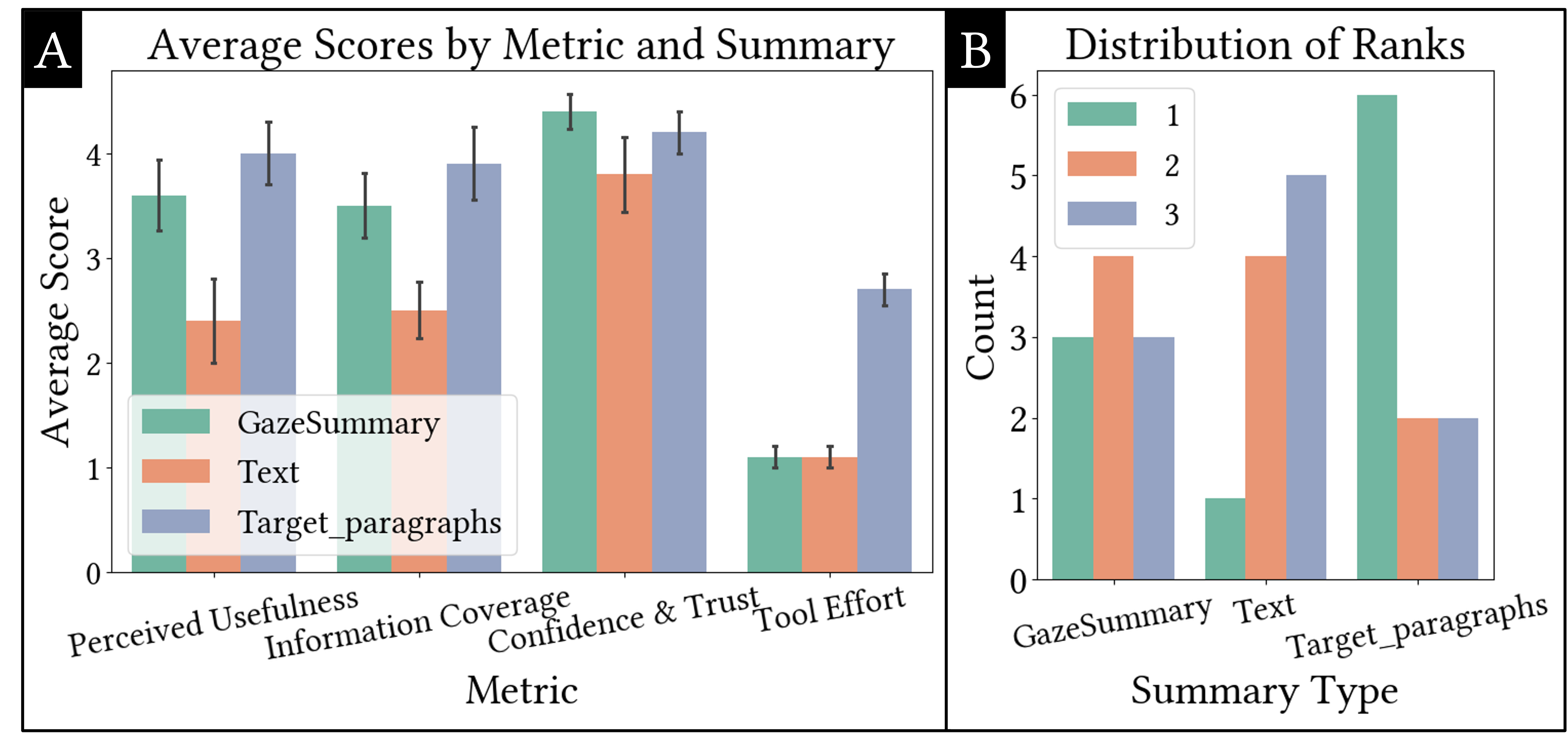}
  \vspace{-0.2in}
  \caption{(A) Personalized summary is more helpful and better aligned with their focus. GazeSummary was easier to use than inputting prompt manually. (B) Explicitly entering prompt increased users’ sense of certainty and led to higher rankings. However, GazeSummary still outperformed the text-only baseline.}
  \label{fig:user_rating_ranking}
  \vspace{-0.2in}
\end{figure}

Fig.~\ref{fig:user_rating_ranking}(A)  shows that both GazeSummary and Target Paragraphs scored significantly higher than Text for \textbf{perceived usefulness}. This indicates that personalized summaries better support users in completing tasks compared to summaries with evenly distributed content. Results for \textbf{information coverage} were similar. Participants generally felt that GazeSummary and Target Paragraphs better reflected the information they attended to. No significant differences were observed across the three methods for \textbf{confidence and trust}, likely because modern LLMs rarely generate factual errors. GazeSummary and Target Paragraphs scored slightly higher, reflecting that summaries aligned with users' attention gave a stronger sense of certainty. Although GazeSummary and Target Paragraphs performed similarly on these three metrics, users reported that Target Paragraphs required more \textbf{time and mental effort} to use. Despite these benefits, most participants ranked Target Paragraphs as first in overall performance, as explicit prompts gave them a stronger sense of control over how the summary aligned with their needs.


\vspace{0.05in}
\noindent\fbox{\parbox[][][]{0.98\linewidth}{\textbf{Takeaway 2:} \textit{Implicit Gaze-based LLM summary is perceived as more useful and has better coverage compared to only text-based LLM summary while requiring significantly lower effort compared to explicit inputs which provide similar benefits.}}}

\section{Discussion and Future Work}

\subsection{Insights on Gaze-Based Personalization}



Our user-centric study of implicit gaze inputs for personalization of text-based LLM tasks demonstrate three key insights:

\noindent \textbf{(1)} We show that implicit gaze inputs can guide LLM to achieve similar personalization as user-provided summary.

\noindent \textbf{(2)} Heatmap-based gaze representations allow LLMs to achieve the best personalization, especially at the phrase level, surpassing explicit-input approaches.

\noindent \textbf{(3)} Our user study in academic reading compares different LLM summary generation methods and shows that gaze-based summaries improve perceived usefulness and focused content coverage compared to text-only summaries, and require less effort to use.




\subsection{Limitations}
Our preliminary design of gaze-based implicit inputs for text-based LLM tasks faced several limitations that need to be addressed in future: \textbf{(1)} No affordable commodity smart glasses come with in-built gaze tracking today. We instead had to rely on static gaze trackers placed on the laptop screen to run the user-study. However, next-gen glasses such as Project Aria do have these gaze-trackers when available. 
\textbf{(2)} An interesting dilemma arose during user grading: the trade-off between attention-focused detail and comprehensiveness.
We believe that a larger-scale user study may explore these trade-offs to derive new metrics of summary personalization. 
\textbf{(3)} Our study focused on text-only tasks while real world lectures/presentations also have images and graphs. Future work can explore extending this gaze-based approach for generalized LLM tasks where user intent matters. (4) Our method may fail in the following cases: First, the insufficient accuracy of wearable eye trackers and drift after prolonged use without recalibration may lead to problems in distinguishing lines of text. Second, it cannot detect the user's mental states and may fail if the user's gaze ray accidentally intersects irrelevant objects during thinking or mind wandering. However, eye-tracking features have the potential to detect cognitive states, which has the potential to address this issue in the future.




\begin{figure}[t]
  \centering
  \includegraphics[width=1.0\linewidth]{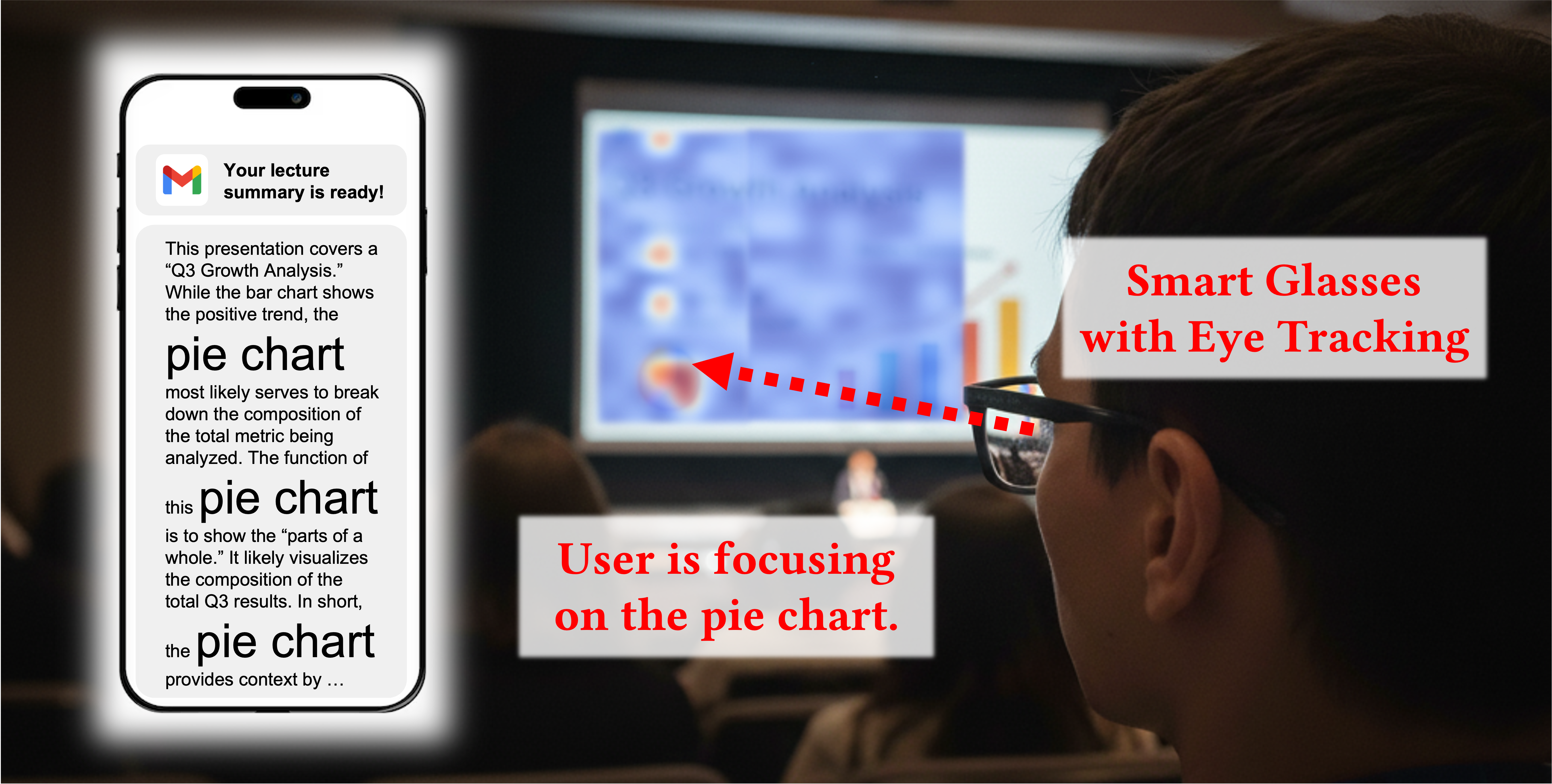}
  \caption{Future work: automatic lecture summary generation based on attention}
  \label{fig:app_lecture}
\end{figure}

\subsection{Future Vision and Next Steps}
We believe that gaze, as a powerful tool, can augment personalized experiences by guiding LLMs as an implicit prompt.
We envision a future where gaze goes beyond a fancy sensing technology but becomes truly integrated in everyday wearables. We envision the following next steps to make a gaze-aware mobile future built on the above study: \textbf{{(1)}} Addressing privacy concerns of capturing gaze by developing gaze-based SLMs on the edge. \textbf{{(2)}} Developing gaze foundational models for co-modal training of LLMs. \textbf{{(3)}} Detecting users’ cognitive and affective states from gaze, such as confusion and engagement, to support adaptive interaction with LLMs.



\bibliographystyle{ACM-Reference-Format}
\bibliography{ref-minimized}

\clearpage
\appendix

\section{Prompts for Different Summarization Methods in Evaluation}
We incorporate definitions of a 'good summary' into the prompt, categorized into comprehensiveness and quality requirements. Additionally, for personalized tasks, a definition of a 'good personalized summary' is included. The corresponding texts are as follows:
\begin{itemize}
    \item Comprehensiveness Requirement: \begin{quote}
        "For comprehensiveness, while covering as much of the user's focused topic as possible, a good summary should also touch on other aspects."
    \end{quote}
    \item Quality Requirement: \begin{quote}
        "For quality, please consider four aspects: (1) Consistency - the factual alignment between the summary and the summarized source. (2) Coherence - the collective quality of all sentences. The summary should be well-structured and well-organized. (3) Relevance - selection of important content from the source. (4) Fluency - the quality of individual sentences."
    \end{quote}
    \item Personalization Requirement: \begin{quote}
        "A good personalized summary should include more contents that the user spend more time reading and touch other contents briefly. For the content that the user is focused on, a better personalized summary should contain more details and be more consistent with the original statements."
    \end{quote}
\end{itemize}

\subsection{Gaze Density}
The gaze density of each sentence is obtained by dividing the total gaze duration by the sentence length. When generating the summary, the top 20\% of sentences are selected and inserted into the prompt. The completed prompt is as follows:

\begin{quote}
    "Generate a one-paragraph 150-word personalized summary for this article based on user's focused sentences. \textit{[Personalization Requirement]} More generally, a good summary should be more comprehensive and of better quality. \textit{[Comprehensiveness Requirement]} \textit{[Quality Requirement]} Do not explain or say any of your analysis. The user spends more time on the following sentences in the article: \textit{[top-attended sentences]} \textbackslash n Article Text: \textit{[source text]}"
\end{quote}

\subsection{Gaze Heatmap}
In this method, gaze heatmaps are generated by spatially aggregating gaze points over the
reading area and visualizing them as overlaid attention maps on the text. The heatmaps are added to the prompt when generating the summary:

\begin{quote}
    "\textit{[Heatmap images]} \textbackslash n Generate a one-paragraph 150-word personalized summary for the following article based on user's gaze heatmap. Bright and warm colors like red or orange mean the user spends more time on it, and dim and cold colors like blue mean the user spends less time on it. \textit{[Personalization Requirement]} More generally, a good summary should be more comprehensive and of better quality. \textit{[Comprehensiveness Requirement]} \textit{[Quality Requirement]} Based on this heatmap and text, you should first identify which sentences and paragraphs the user spends more time reading and then generate a personalized summary of this article based on these contents. Do not explain or say any of your analysis. \textbackslash n Article Text: \textit{[source text]}"
\end{quote}

\begin{figure}
    \centering
    \includegraphics[width=1.0\linewidth]{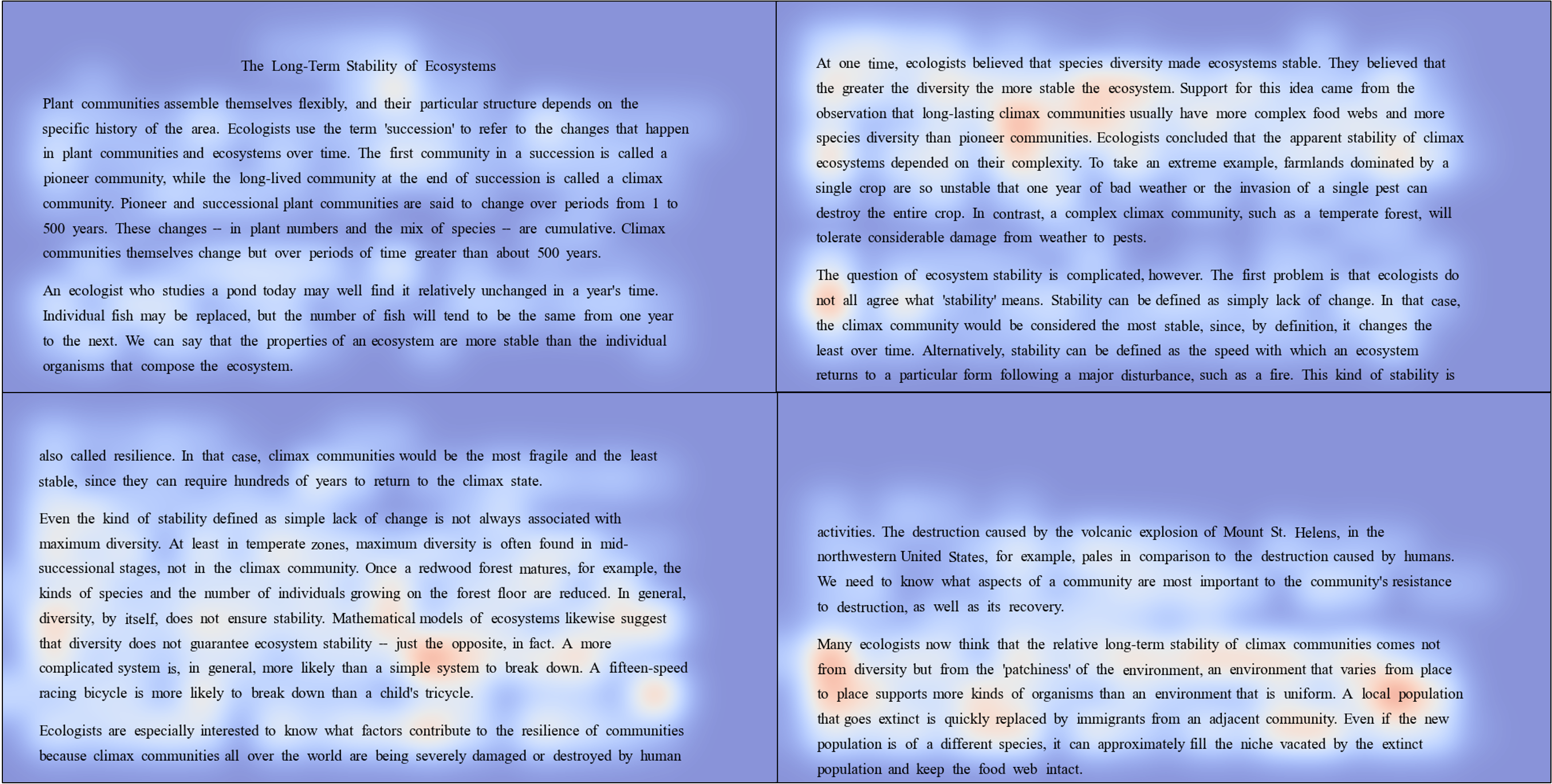}
    \caption{An example of heatmaps of one article.}
    \label{fig:heatmap_example}
\end{figure}

\subsection{Gaze Attention SVM}

We identified 26 gaze features for attention level classification, following the methodology of \cite{bixler_automatic_2016}. To ensure robustness across varying experimental layouts, we excluded absolute distance-based metrics, such as saccade distance. The final feature set comprises:
\begin{enumerate}
    \item Fixation duration: Duration in milliseconds of a fixation.
    \item Saccade duration: Duration in milliseconds between two subsequent fixations.
    \item Pupil diameter: Diameter of pupil (standardized within-participant).
    \item Number of saccades: Total number of saccades within window.
    \item Fixation saccade ratio: Ratio of fixation duration to saccade duration.
\end{enumerate}
For the first three categories, we calculated eight descriptive statistics, which are minimum, maximum, mean, median, standard deviation, skewness, kurtosis, and range, resulting in 24 derived features. Combined with saccade count and the fixation-saccade ratio, this yielded a 26-feature vector. Based on these features, sentences ranked in the top 20\% for predicted attention are integrated into the prompt. The prompt structure mirrors the gaze density condition but is subtly adjusted to emphasize classifier-driven attention estimation:

\begin{quote}
   "Generate a one-paragraph 150-word personalized summary for this article based on user's focused sentences. \textit{[Personalization Requirement]} More generally, a good summary should be more comprehensive and of better quality. \textit{[Comprehensiveness Requirement]} \textit{[Quality Requirement]} Do not explain or say any of your analysis. Sentences below are classified as focused sentences which means the user spent more time reading these sentences: \textit{[top-attended sentences]} \textbackslash n Article Text: \textit{[source text]}" 
\end{quote}

\subsection{Target Paragraphs}
This baseline simulates an explicit prompting scenario where users manually specify sections for inclusion to ensure the summary covers their specific information needs. Specifically, the indices of the target paragraphs assigned to participants prior to the reading session are integrated into the prompt. This process biases the LLM toward the user's pre-defined focus, resulting in a personalized summary based on explicit intent:

\begin{quote}
    "Generate a one-paragraph 150-word personalized summary for this article based on user's focus. \textit{[Personalization Requirement]} More generally, a good summary should be more comprehensive and of better quality. \textit{[Comprehensiveness Requirement]} \textit{[Quality Requirement]} Do not explain or say any of your analysis. The \textit{[target paragraph id1]} and \textit{[target paragraph id2]} paragraphs are the ones that users pay most attention to. \textbackslash n Article Text: \textit{[source text]}"
\end{quote}

\subsection{Text-Only Method}
This baseline generates a generic summary lacking personalization. The prompt consists solely of the source text without any indicators of user preferences or attention weights:

\begin{quote}
    "Generate a one-paragraph 150-word summary for this article. Generally, a good summary should be more comprehensive and of better quality. \textit{[Comprehensiveness Requirement]} \textit{[Quality Requirement]} The original article is shown below. Do not explain or say any of your analysis. \textbackslash n Article Text: \textit{[source text]}"
\end{quote}

\section{Case Study Materials and System Implementation}
\subsection{Reading Materials}
We selected three reading materials from the TOEFL test, covering distinct disciplines: Biology (Symbiotic Relationships\footnote{https://top.zhan.com/toefl/read/practicereview-275-13.html}), Art (The Origins of Theater\footnote{https://top.zhan.com/toefl/read/practicereview-2-13.html}), and Geography (Petroleum Resources\footnote{https://top.zhan.com/toefl/read/practicereview-54-13.html}). Each article consists of approximately 700 words distributed across 5–6 paragraphs.

\subsection{Task Description}
The tasks specifically associated with the two paragraphs presented to users before each reading session are detailed below. These tasks are designed to direct the users' attention toward specific paragraphs, thereby emulating real-world scenarios where readers approach a text with predefined objectives.

Biology:
\begin{quote}
    "You are a biology student preparing for an exam. Your professor has assigned you a 3-bullet point study guide outline. Your outline must only cover the topic of \textbf{Parasitism}.
    
    You are instructed to \textbf{ignore the other symbiotic types} (commensalism and mutualism). Your outline must focus only on \textbf{parasitism} and \textbf{a case study} about rabbits related to how natural selection shapes this relationship."
\end{quote}

Art:
\begin{quote}
    "You are a cultural anthropology student. Your professor has assigned you a 3-bullet point outline for a report. Your report must only describe the most widely accepted anthropological theory of theater's origin (the \textbf{Myth-Ritual} theory) and how it evolves into \textbf{autonomous theater}.
    
    You are instructed to \textbf{ignore alternative theories} (like storytelling or dance) and ignore the psychological motives. Your outline must trace the specific steps of \textbf{Myth-Ritual} theory's evolutionary process."
\end{quote}

Geography:
\begin{quote}
    "You are a junior analyst at an energy investment firm. Your manager is preparing a report for investors on the \textbf{difficulty and risks} of oil exploration. She \textbf{does not need} a summary of how oil is formed, the basic drilling process, or its environmental impact.
    
    She wants a 3-bullet point outline for her report that focuses only on the major \textbf{challenges and limitations} that make finding and recovering oil \textbf{difficult and expensive}. Your task is to write this ``Challenges \& Limitations" outline, including modern exploration challenges and oil recovery limitations."
\end{quote}

\subsection{Comprehension Questions}
According to the official TOEFL guidelines\footnote{https://www.ets.org/toefl/test-takers/ibt/about/content/reading.html}, there are six types of reading questions, such as Factual Information, Inference, and Vocabulary. For our study, we selected two factual questions and one inference question related to the target paragraphs that are largely independent of the participants' prior knowledge. This selection ensures that users must engage deeply with the target passages to arrive at the correct answers. The specific questions are listed below:

Biology:
\begin{enumerate}
    \item Which of the following is true of the action of natural selection on hosts and parasites?
    \begin{enumerate}
        \item Hosts benefit more from natural selection than parasites do.
        \item Both aggression in predators and defensive capacities in hosts are favored for species survival.
        \item The ability to make toxic chemicals enables a parasite to find and isolate its host.
        \item Larger size equips a parasite to prey on smaller host organisms.
    \end{enumerate}
    \item Which of the following can be concluded from the discussion about the Australian rabbit population?
    \begin{enumerate}
        \item Human intervention may alter the host, the parasite. and the relationship between them.
        \item The risks of introducing outside organisms into a biological community are not worth the benefits.
        \item Humans should not interfere in host-parasite relationships.
        \item Organisms that survive a parasitic attack do so in spite of the natural selection process.
    \end{enumerate}
    \item All of the following characterize the way natural selection stabilized the Australian rabbit population EXCEPT:
    \begin{enumerate}
        \item The most toxic viruses died with their hosts.
        \item The surviving rabbits were increasingly immune to the virus.
        \item The decline of the mosquito population caused the spread of the virus to decline.
        \item Rabbits with specific genetic make-ups were favored.
    \end{enumerate}
\end{enumerate}

Art:
\begin{enumerate}
    \item Theories of the origins of theater
    \begin{enumerate}
        \item are mainly hypothetical
        \item are well supported by factual evidence
        \item have rarely been agreed upon by anthropologists
        \item were expressed in the early stages of theater's development
    \end{enumerate}
    \item Why did some societies develop and repeat ceremonial actions?
    \begin{enumerate}
        \item To establish a positive connection between the members of the society
        \item To help society members better understand the forces controlling their food supply
        \item To distinguish their beliefs from those of other societies
        \item To increase the society's prosperity
    \end{enumerate}
    \item What may cause societies to abandon certain rites?
    \begin{enumerate}
        \item Emphasizing theater as entertainment
        \item Developing a new understanding of why events occur
        \item Finding a more sophisticated way of representing mythical characters
        \item Moving from a primarily oral tradition to a more written tradition
    \end{enumerate}
\end{enumerate}

Geography:
\begin{enumerate}
    \item Which of the following strategies for oil exploration is described in terms of modern exploration difficulty?
    \begin{enumerate}
        \item Drilling under the ocean's surface
        \item Limiting drilling to accessible locations
        \item Using highly sophisticated drilling equipment
        \item Constructing technologically advanced drilling platforms
    \end{enumerate}
    \item What does the development of the Alaskan oil field demonstrate?
    \begin{enumerate}
        \item More oil is extracted from the sea than from land.
        \item Drilling for oil requires major financial investments.
        \item The global demand for oil has increased over the years.
        \item The North Slope of Alaska has substantial amounts of oil.
    \end{enumerate}
    \item The decision to drill for oil depends on all of the following factors EXCEPT
    \begin{enumerate}
        \item permission to access the area where oil has been found
        \item the availability of sufficient quantities of oil in a pool
        \item the location of the market in relation to the drilling site
        \item the political situation in the region where drilling would occur
    \end{enumerate}
\end{enumerate}

\subsection{Prototype Implementation}
The academic reading assistant prototype was developed using a Python backend and a React frontend. The system handles three data pipelines based on the experimental condition:

\begin{itemize}
    \item GazeSummary (Implicit Prompt): Gaze data is captured by a Tobii Pro Spark eye tracker and recorded in the backend once the user initiates the session via the "Start Recording" button.
    \item Target Paragraphs (Explicit Prompt): For this baseline, user prompts are collected through a dedicated "Input Page" on the frontend after the user concludes the reading session by clicking "End Recording".
    \item Text (Non-personalized): For this baseline, the system retrieves only the original source text without additional attention information.
\end{itemize}

These data streams are transmitted to the Gemini 2.5 Pro model via API to generate the respective summaries. The generated outputs are then returned to the frontend and displayed to the user as cognitive support during their writing tasks.

\begin{figure}
    \centering
    \includegraphics[width=1.0\linewidth]{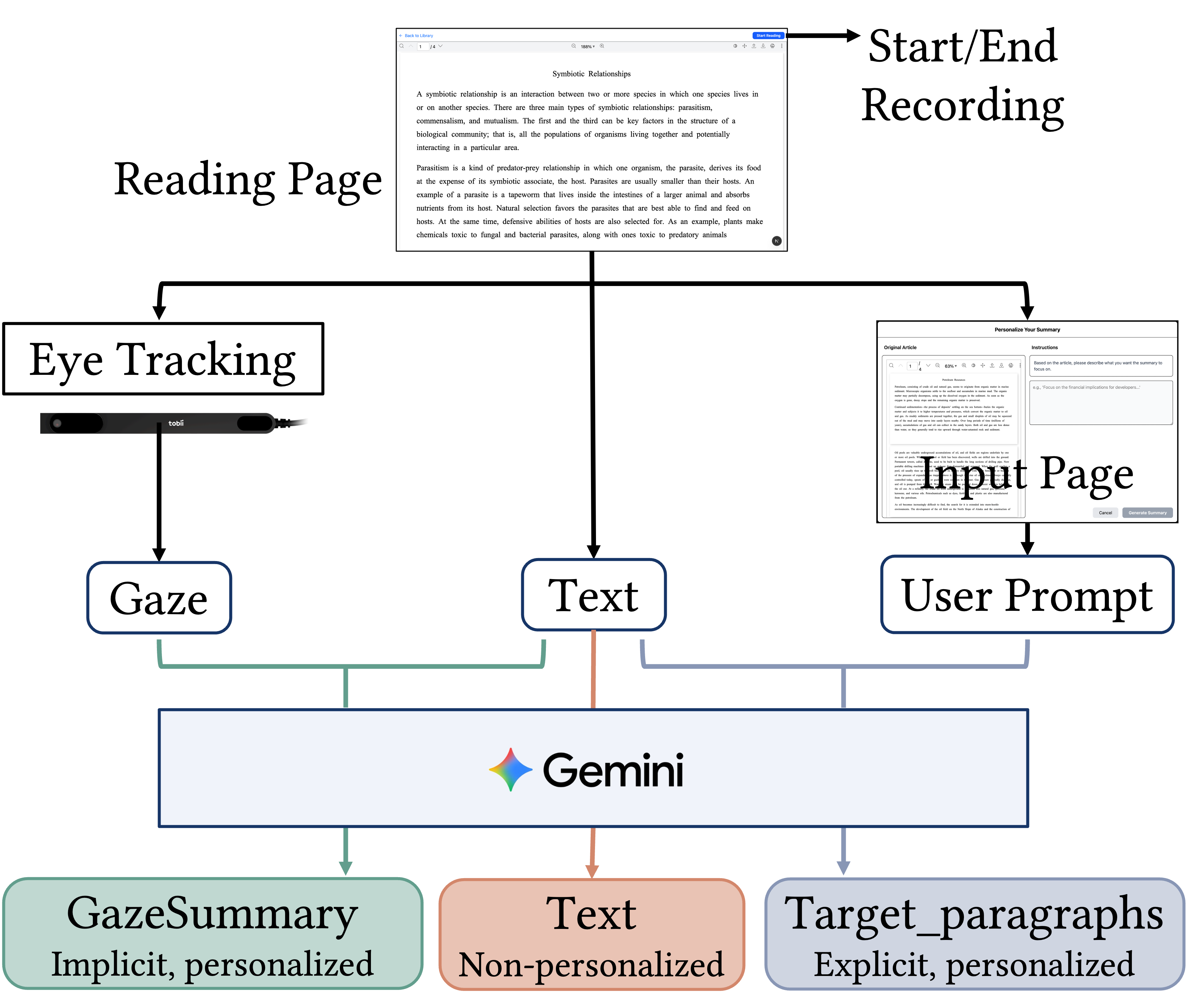}
    \caption{System architecture of the academic reading assistant prototype, illustrating how gaze data, original text, and user prompts are integrated through the Gemini in different experiment settings.}
    \label{fig:prototype}
\end{figure}

\end{document}